\documentclass[12pt]{article}   %% REVTeX 4.0
\usepackage{ol2}
\usepackage[draft]{hyperref}
\usepackage{amsmath}
\begin{document}
%\twocolumn[ %% activate for two-column option

\title{Extreme value statistics in coupled lasers}

%% For REVTeX it is possible to automate superscript and e-mail callouts with the superscriptaddress option; see REVTeX4 documentation.

\author{Moti Fridman, Rami Pugatch, Micha Nixon, Asher A. Friesem, and Nir Davidson$^{*}$}
\address{
Weizmann Institute of Science, Dept. of Physics of Complex Systems,
Rehovot 76100, Israel \\ $^*$Corresponding author:
nir.davidson@weizmann.ac.il }

\begin{abstract}
Experimental configuration for investigating the dynamics and the statistics of the phase locking level of coupled lasers that have no common frequency is presented. The results reveal that the probability distribution of the phase locking level of such coupled lasers fits a Gumbel distribution that describes the extreme value statistic of Gaussian processes. A simple model, based on the spectral response of the coupled lasers, is also described, and the calculated results are in good agreement with the experimental results.
\end{abstract}

% REPLACE WITH CORRECT OCIS CODES FOR YOUR ARTICLE
                          % NOTE: \ocis{} IS ALIASED TO \pacs{} BUT MUST
                          % FORMAT THE TERMS CORRECTLY FOR EACH JOURNAL

\maketitle %% null function with osajnl.sty
%]% %% activate for two-column option

\noindent

Phase locking of coupled oscillators was studied over the years in many different contexts including chemical oscillators with mutual coherence~\cite{V1}, arrays of Josephson junctions that are frequency locked~\cite{V2} and arrays of coupled lasers that are phase locked~\cite{V3, V4, V5}. In these, complete phase locking occurs when all the oscillators have at least one common frequency. When there is no common frequency, the oscillators group in several clusters, where each cluster oscillates at a different frequency~\cite{Strogatz}. So far there is very little, if any, experimental investigations which deal with the dynamics and statistics of coupled oscillators that do not have a common frequency.

In this letter, we deal with the phase locking level in an array of lasers that have no common frequency, and show that the distribution of the phase locking level is in good agreement with the Gumbel distribution function which describes the extreme value statistics from Gaussian processes~\cite{Gumbel1, Gumbel2}. Specifically, we investigate the phase locking level of an array of 25 coupled fiber lasers. Although each fiber laser support 100,000 eigenfrequencies, the probability to find a common frequency for all the lasers in the array is very small ($<10^{-5}$)~\cite{no_more_shirakawa, no_more_rothenberg, no_more_shakir}, so the lasers group in several clusters, each with its own frequency~\cite{Galvanauskas16, Moti25}. Due to thermal and acoustic fluctuations, the length of each fiber laser and its corresponding eigenfrequencies changes rapidly and randomly. As we show below, phase locking minimizes loss in the array, so mode competition will favor frequencies that maximize the size of the phase locked clusters at each moment~\cite{VarditPRL}. Since the distribution of the phase locking level for different frequencies is Gaussian, the statistics of the maximum phase locking level is described by the Gumbel distribution function.

\begin{figure}[h]
\centerline{\includegraphics[width=8cm]{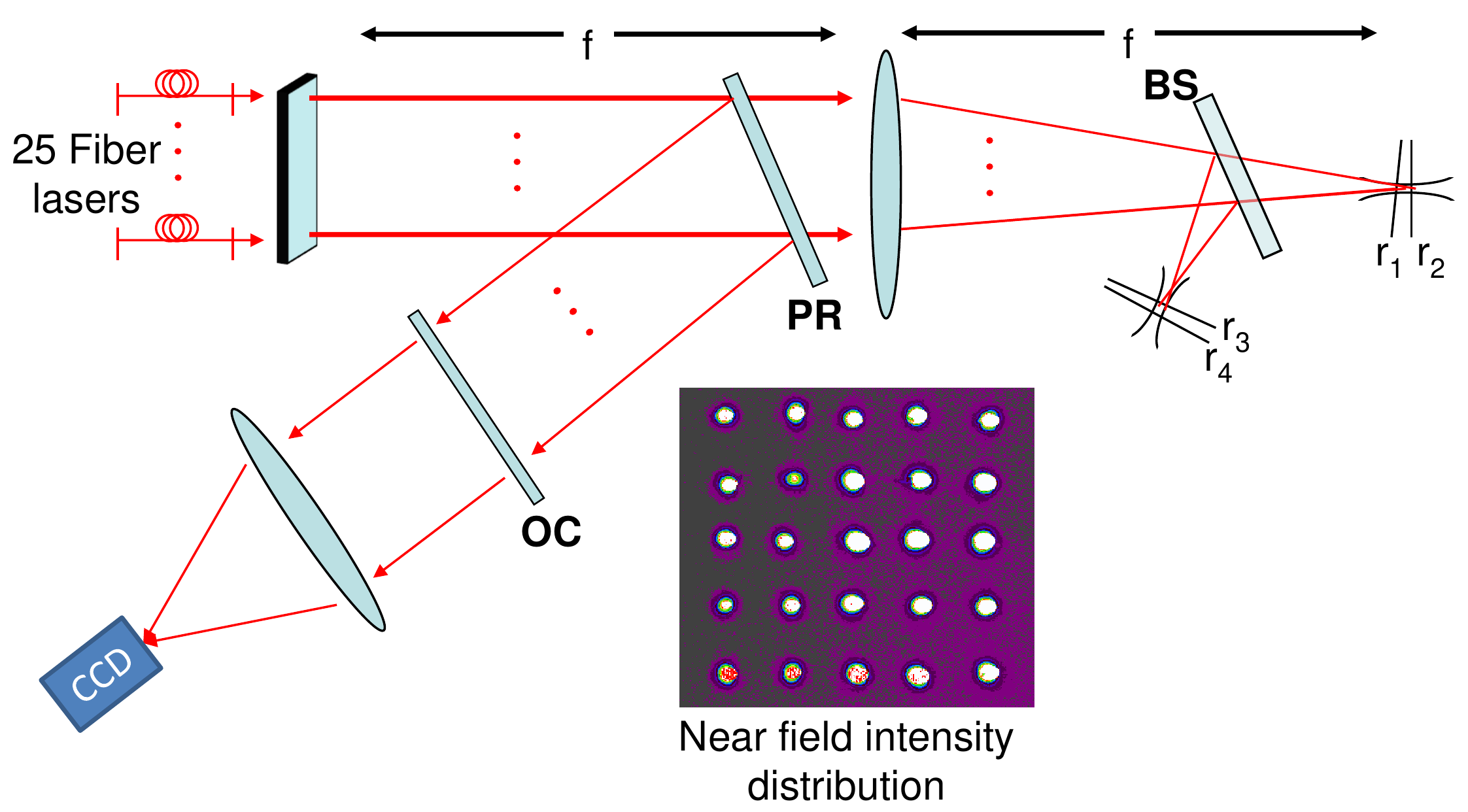}}
\caption{\label{setup}Experimental configuration for phase locking an array of fiber lasers and for determining their phase locking level. OC - output coupler, PR - partial reflector, BS - 50\% beam splitter. Inset - near field intensity distribution when all 25 fiber lasers are operating.}
\end{figure}

The experimental configuration that we used for measuring the phase locking level of an array of fiber lasers is presented in Fig.~\ref{setup} and described in detail in~\cite{Moti25}. Briefly, each fiber laser was comprised of a Ytterbium doped fiber, a rear high reflection ($>99\%$) fiber Bragg grating (FBG), and a front low reflecting ($5\%$) FBG both with a $10nm$ bandwidth. Each laser was pumped through the rear FBG with a $975nm$ diode laser at $200mW$, and after the front FBG we attached a collimator to obtain a $0.4mm$ diameter beam. The collimators of all the 25 lasers were accurately aligned in a $5X5$ square array of parallel beams with parallelism better than $0.1mrad$. The separation between adjacent beams was $3.6mm$. A representative near-field intensity distribution, measured close to the output coupler when all 25 fiber lasers are operating, is presented at the lower inset of Fig.~\ref{setup}. We determined the length of each fiber laser by measuring the longitudinal mode beating frequency at their output by means of a fast photodetector connected to a RF spectrum analyzer. We found that the distribution of the lasers lengths is Gaussian with a mean value of $3m$ and a width of $0.5m$. The intensity of each fiber laser was about $100mW$ which is much above threshold but still low enough so nonlinear effects where negligible~\cite{Corcoran}.

The coupling between the fiber lasers was achieved by means of four coupling mirrors denoted as $r_1$, $r_2$, $r_3$ and $r_4$ with reflectivity of $40\%$ for $r_1$ and $r_3$ and reflectivity of $100\%$ for $r_2$ and $r_4$. All the coupling mirrors were located close to the focal plane of a focusing lens with $500mm$ focal length, forming a self imaging cavity with the array. Since there was only enough space for one pair of mirrors within the Rayleigh range of the focussing lens, we inserted a $50\%$ beam splitter to obtain another focal plane where we placed another pair of mirrors. By controlling the orientations of the coupling mirrors we could realize a variety of connectivities for the fiber lasers in the array, and in our experiments we concentrated on the two-dimensional nearest neighbors connectivity~\cite{Moti25}. Finally, we directed about $10\%$ of the light with a partially reflecting mirror (PR), towards an output coupler (OC) of 99\% reflectivity. The OC was placed at a distance of 2f from the collimator array, and reflected part of the light from each laser back onto itself with the same delay as the light that is coupled from the other lasers~\cite{michaTime}.

%Such an arrangement ensures that although the coherence length of each laser is shorter than the  distance to the coupling mirrors, phase locking can still occur with a uniform phase across the array~\cite{michaTime}.

\begin{figure}[h]
\centerline{\includegraphics[width=8cm]{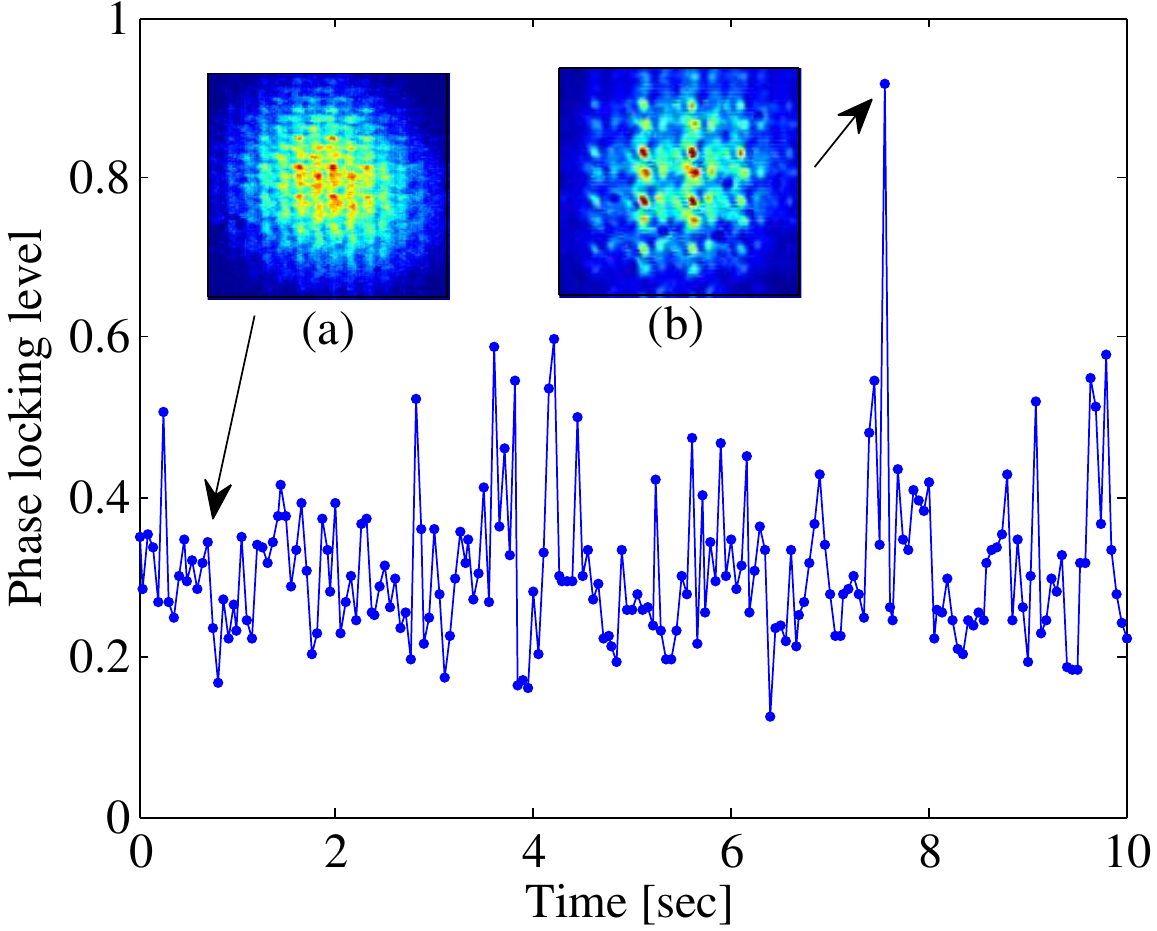}}
\caption{\label{bareData} Typical experimental results of the phase locking level as a function of time over a 10 seconds interval. The phase locking level was determined from the far field intensity distribution of the output. Insets show typical far field intensity distributions - (a) low fringe visibility where the phase locking level is low; (b) high fringe visibility were the phase locking level is high.}
\end{figure}

We measured the phase locking level as a function of time for different number of lasers in the array. This was done by continuously detecting the far-field intensity distribution of the interference pattern of all the light from the array with a CCD camera, determining the maxima and minima intensities, and calculating the average fringe visibility along the x and y directions. The fringe visibility provides a direct measure for the phase locking level that ranges from 0 to 1. The correlation time of the phase locking level is shorter than $100msec$, so over a 10 hours period we acquired about $370,000$ uncorrelated measurements of the fringe visibility. Representative experimental results of the fringe visibility as a function of time for 10 seconds interval are presented in Fig.~\ref{bareData}. The insets show two typical far field intensity distributions - one with low fringe visibility and the other with high fringe visibility.

\begin{figure}[h]
\centerline{\includegraphics[width=8cm]{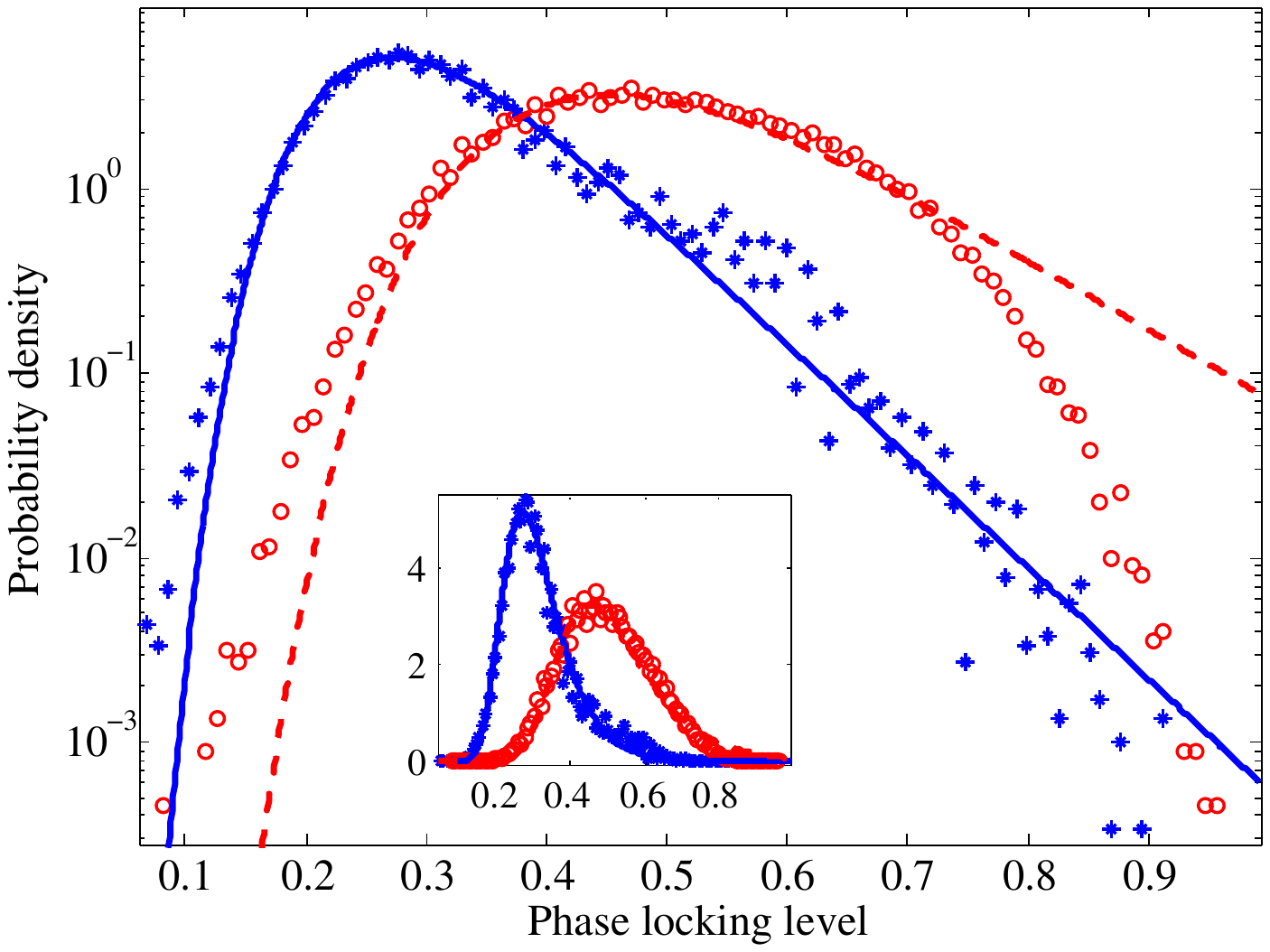}}
\caption{\label{gumbel} Probability distributions of the phase locking level for two arrays of fiber lasers. Asterisks (blue) - experimental probability distribution of the measured phase locking level of an array with 25 fiber lasers; circles (red) - experimental probability distribution of the measured phase locking level of an array with 12 fiber lasers; solid and dashed curves - Gumbel distributions fitted for both cases. Inset shows the probability distributions with the Gumbel fits in linear scale.}
\end{figure}

Using the 370,000 measurements, we determined the probability distribution of the fringe visibility and fitted it to the distribution of extreme values in Gaussian processes, namely the BHP distribution~\cite{BHP}. A simplified form of BHP distribution is given by
\begin{equation}
P(x)=e^{c \left( \frac{x-\alpha}{\beta}-e^{\frac{x-\alpha}{\beta}} \right)}, \label{Eq1}
\end{equation}
where $\alpha$ denote the mean value, $\beta$ the width and $c$ the measure for correlations in the Gaussian process, with $c=1.58$ indicating highly correlated process and $c=1$ an uncorrelated process. After fitting our probability distribution to the BHP distribution, we determined that $c=1.03$, indicating that the Gaussian process in our case is uncorrelated. Consequently, the BHP distribution of Eq.~(\ref{Eq1}), reduces to the generalized Gumbel distribution~\cite{Gumbel1}. Representative experimental probability distributions for an array of 25 fiber lasers and an array of 12 fiber lasers with fits to a generalized Gumbel distribution functions are presented in Fig.~\ref{gumbel}. The inset shows the corresponding results in a linear scale. As evident, there is a very good agreement between the experimental probability distributions of the phase locking level and the Gumbel distribution function for the array of 25 fiber lasers, but not as good for the array of 12 fiber lasers.

First, we explain qualitatively the connection between the phase locking level and the Gumbel distribution. The number of lasers in each phase locked cluster changes rapidly and randomly, but always while maximizing the phase locking level in the array. This maximum phase locking level occurs at a specific frequency out of all the possible frequencies within the FBG bandwidth where the lasers losses are minimal. By considering the spectral response of the coupled lasers it was shown, that the distribution of the phase locking levels for all the available frequencies is an uncorrelated Gaussian distribution~\cite{motiCommonLM, no_more_shakir}. Since the distributions of maxima of uncorrelated Gaussian processes are described by the Gumbel distribution functions~\cite{Gumbel1, Gumbel2}, the probability distribution of the phase locking level should be the same.

As evident from the results of Fig.~\ref{gumbel}, for the array of 25 fiber lasers the average phase locking level is 0.28. Accordingly, rare events with more than triple the average phase locking level can occur, so there is good agreement between the experimental probability distribution and the Gumbel distribution for phase locking levels up to 0.85. This is not the case for smaller arrays of fiber lasers where the average phase locking level is higher and the effect of clipping of the distribution at unity phase locking level is significant. To illustrate this effect, we measured the phase locking level for smaller arrays, with 20, 16 and 12 fiber lasers, and found that as the number of fiber lasers in the array decreases, the fit for a Gumbel distribution is less exact. This is clearly evident for the array of 12 fiber lasers where the average phase locking level is 0.57 and the agreement to the Gumbel distribution is only good for phase locking level below 0.7.

Next, we present a simple quantitative model that relates the phase locking level to an extreme value in the spectral response of the array of coupled lasers. We start by assuming no gain and determining the spectral response of each laser cavity when we replaced all components that couple light into the laser cavity with an effective front mirror. The reflectivity of this effective mirror depends on the frequency~\cite{motiCommonLM}. For example, the effective reflectivity of the $i$'th laser in a one-dimensional array of N coupled lasers, when each of which is coupled to its two nearest neighbors, is
\begin{equation}
R_i=\frac{1}{1-r(1-2 \kappa)+\frac{\kappa^2}{1-R_{i-1}^{(u)} e^{2 \imath k l_{i-1}}}+\frac{\kappa^2}{1-R_{i+1}^{(d)} e^{2 \imath k l_{i+1}}}} \label{Eq2}
\end{equation}
where $\kappa$ denote the coupling to the two neighbors, $r$ the reflectivity of the output coupler, $l_i$ the length of the $i$'th laser, $k$ the propagation vector, and $R_i^{(u)}$ and $R_i^{(d)}$ the effective reflectivities from all the lasers above and below the $i$'th laser, given by
\begin{equation}
R_i^{(u)}=\frac{1}{1-r(1-2 \kappa)+\frac{\kappa^2}{1-R_{i-1}^{(u)} e^{2 \imath k l_{i-1}}}},
\end{equation}
and
\begin{equation}
R_i^{(d)}=\frac{1}{1-r(1-2 \kappa)+\frac{\kappa^2}{1-R_{i+1}^{(d)} e^{2 \imath k l_{i+1}}}}, \label{Eq4}
\end{equation}
and $R_1^{(u)}=R_N^{(d)}=0$. Now, introducing the laser gain together with a mode competition results in amplifications only at frequencies with high effective reflectivity.

\begin{figure}[h]
\centerline{\includegraphics[width=8cm]{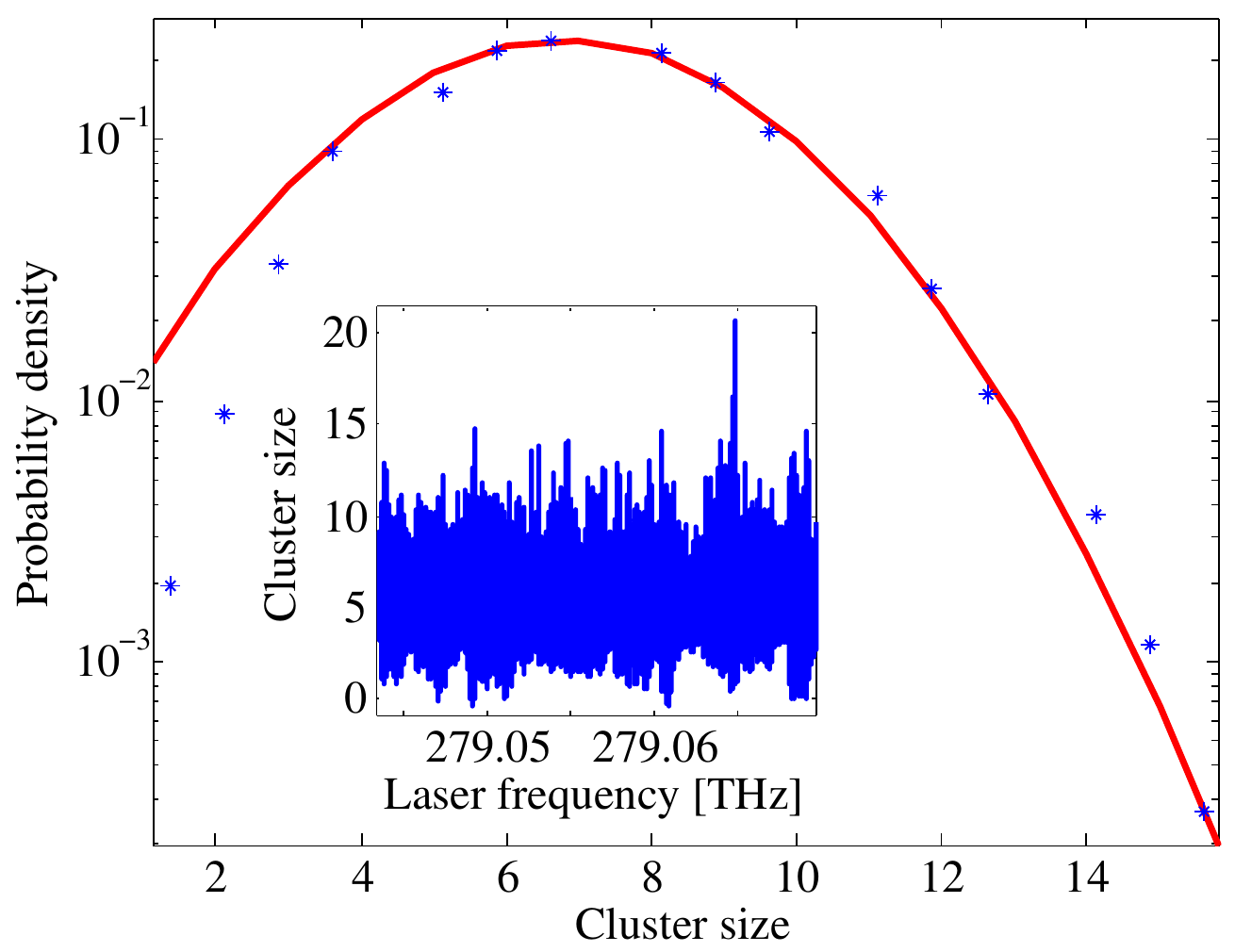}}
\caption{\label{cluster_size} Calculated probability distribution of the size of the main cluster together with a Gaussian fit for clusters larger than the mean size. Inset shows representative calculated results of the main cluster size as a function of the laser frequency.}
\end{figure}

We solved Eqs.~(\ref{Eq2})-(\ref{Eq4}) explicitly for an array of 25 fiber lasers and obtained the effective reflectivity of each laser as a function of the frequency. By counting the number of lasers with effective reflectivity above a certain threshold, we obtained the number of lasers in the main cluster as a function of frequency which provided a direct measure for the phase locking level of the array as a function of frequency~\cite{BHP}. Representative results are shown in Fig.~\ref{cluster_size}. It shows the probability distribution of the size of the main cluster together with a Gaussian fit for the tail of the distribution. The inset shows the size of the main cluster as a function of the frequency. As evident, the calculated probability distribution above the mean size of the main cluster is Gaussian. Then we selected the maximum size of the main cluster within the bandwidth of the FBG ($10nm$) and determined the resulting phase locking level as the ratio of the size of the main cluster over the size of the array.

\begin{figure}[h]
\centerline{\includegraphics[width=8cm]{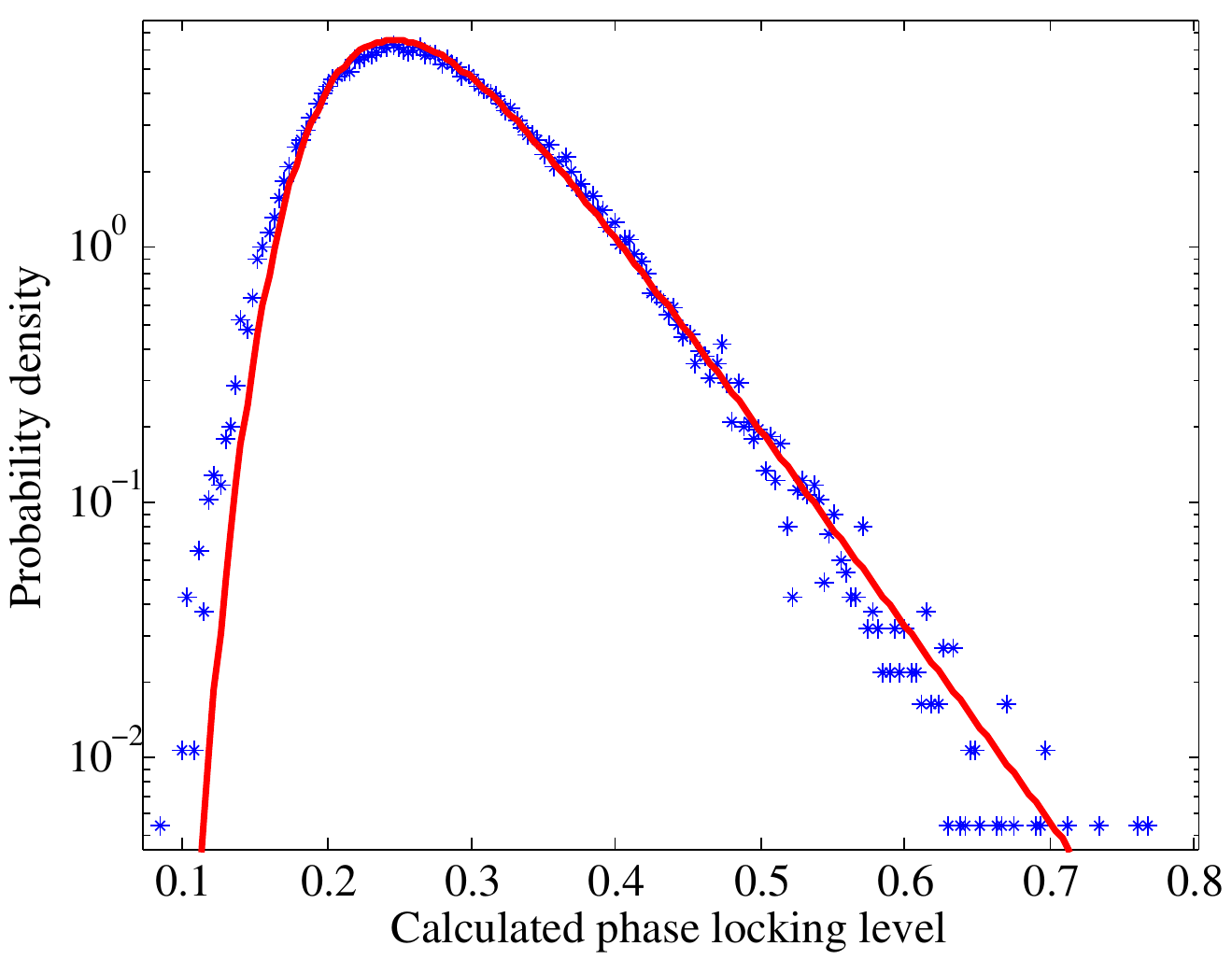}}
\caption{\label{num_eff_ref} Calculated probability distribution of the phase locking level from 50,000 different arrays of fiber lasers as well as a fit to a Gumbel distribution.}
\end{figure}

We repeated these calculations for 50,000 different arrays each with a different random fibers lengths. The fiber length of the $i$'th laser in each array was chosen to be $l_i+\Delta l_i$, where $l_i$ is the measured length of the $i$'th fiber and $\Delta l_i$ a random length taken from a normal distribution with 0 mean and $10\mu m$ width. The results are presented in Fig.~\ref{num_eff_ref}. It shows the probability distribution of the calculated phase locking level together with a fit to a Gumbel distribution. As evident, there is a very good agreement between the distribution of the calculated results and the Gumbel distribution, indicating that the effective reflectivity is suitable for modeling arrays of coupled lasers, and that the underlaying Gaussian process is the number of lasers in the main cluster.

To conclude, we measured the probability distributions of the phase locking level for arrays of fiber lasers, and showed that they fit to Gumbel distributions. We also presented that the underlying Gaussian process is the number of lasers that have a common frequency. Our results can now be exploited to predict for arrays with an arbitrary number of fiber lasers what will be the probability to obtain a specific phase locking level. Finally, Finally,
by operating the fiber lasers close to threshold we observe strong fluctuations also in the total power of the array which can be related to the statistics of extreme eigen values of random matrices~\cite{TWM}.

%\begin{acknowledgments}

This research was supported by the Israeli Ministry of Science and Technology and by the USA-Israel Binational Science Foundation.

%\end{acknowledgments}


\begin{thebibliography}{99}

\bibitem{V1} I. Z. Kiss, Y. Zhai, and J. L. Hudson, Science 296, 1676 (2002).
\bibitem{V2} K. Wiesenfeld, Physica (Amsterdam) 222B, 315 (1996).
\bibitem{V3} I. Kanter, N. Gross,E. Klein, E. Kopelowitz, P. Yoskovits, L. Khaykovich, W. Kinzel, and M. Rosenbluh, Phys. Rev. Lett. 98, 154101 (2007).
\bibitem{V4} G. D. VanWiggeren and R. Roy, Science 279, 1198 (1998).
\bibitem{V5} G.V. Osipov, B. Hu, C. Zhou, M.V. Ivanchenko, and J. Kurths, Phys. Rev. Lett. 91, 024101 (2003).
\bibitem{Strogatz} S.~H.~Strogatz, \newblock {\em Nature} {\bf 410}, 268 (2001).
\bibitem{Gumbel1} E. J. Gumbel, {\em Statistics of Extremes.} Mineola, NY: Dover, 2004.
\bibitem{Gumbel2} N.~Johnson, S.~Kotz, and N.~Balakrishnan, {\em Continuous Univariate Distributions,} Vol. 2, 2nd ed. New York: Wiley, 1995.
\bibitem{no_more_shakir} E.~J.~Bochove, S.~A.~Shakir, "Analysis of a Spatial-Filtering Passive Fiber Laser Beam Combining System" {\em IEEE J. of Selected Topics in Quantum Elec.} {\bf 15}, 320 (2009).
\bibitem{no_more_shirakawa}A.~Shirakawa, K.~Matsuo, and K.~Ueda, "Fiber Laser Coherent Array for Power Scaling of Single-Mode Fiber Laser", \newblock {\em Proc. SPIE}, {\bf 5662}, 482, (2004).
\bibitem{no_more_rothenberg}J.~E.~Rothenberg, "Advances in Fiber Laser Beam Combination" in \newblock {\em Proc. SPIE} {\bf 6873}, 687315 (2008).
\bibitem{Moti25} M.~Fridman, M.~Nixon, N.~Davidson, and A.~A.~Friesem "Passive phase locking of 25 fiber lasers" {\em Opt. Lett.} {\bf 35,} 1434 (2010).
\bibitem{Galvanauskas16} W.~Chang, T.~Wu, H.~G.~Winful, and A.~Galvanauskas, "Array size scalability of passively coherently phased fiber laser arrays," {\em Opt. Express} {\bf 18,} 9634, (2010).
\bibitem{VarditPRL} V.~Eckhouse, M.~Fridman, N.~Davidson, and A.~A.~Friesem, \emph{Phys. Rev. Lett.} {\bf 100}, 024102, (2008).
\bibitem{Corcoran} C.~J.~Corcoran and F.~Durville "Passive Phasing in a Coherent Laser Array" {\em IEEE J.  Quantum Electron.} {\bf 15,}  294 (2009).
\bibitem{michaTime} M.~Nixon, M.~Fridman, E.~Ronen, A.~A.~Friesem, and N.~Davidson, , \newblock {\em Opt. Lett.} {\bf 34}, 1864 (2009).
\bibitem{BHP} S.~T.~Bramwell, K.~Christensen, J.~Y.~Fortin, P.~C.~W.~Holdsworth, H.~J.~Jensen, S.~Lise, J.~Lopez, M.~Nicodemi, J.~F.~Pinton, and M.~Sellitto,"Universal Fluctuations in Correlated Systems" {\em Phys. Rev. Lett.}, {\bf 84}, 3744, (2000).
\bibitem{motiCommonLM} M.~Fridman, M.~Nixon, E.~Ronen, A.~A.~Friesem, and N.~Davidson "Phase locking of two coupled lasers with many longitudinal modes" \newblock {\em Opt. Lett.} {\bf 35}, 526 (2010).
\bibitem{TWM} M.~Fridman, R.~Pugatch, M.~Nixon, A.~A.~Friesem, and N.~Davidson, "Measuring maximal eigenvalue distribution of Wishart random matrices with coupled lasers" {\em Submitted to PRL}.
\end{thebibliography}
\end{document}